\documentclass{amsart} 
\usepackage{varioref,graphicx,booktabs}

\title{Modularity and heavy-tailed degree distributions}
\author{Larry Wilson}
\address{Center for Communications Research--La Jolla}
\email{larry@ccrwest.org}

\begin{document}
\begin{abstract} 
 Identifying clusters of vertices in graphs continues to be
an important problem and \textit{modularity} continues to be used as
a tool for solving the problem.  Modularity, which measures the
quality of a division of the vertices into clusters, explicitly treats
vertices of different degrees differently, imposing a larger penalty
when high-degree vertices are put in the same cluster. We claim that
this unequal treatment negatively impacts the performance of
clustering algorithms based on modularity for graphs with heavy-tailed degree
distributions. We used the Greedy Modularity hill-climb
to find clusters in graphs with power-law degree
distributions and observed that it performed poorly clustering
low-degree vertices. We propose a simple variant of modularity that we
call \textit{flat modularity}. We found that using the same algorithm
with the modified score instead improved the performance of the
clustering algorithm on low-degree vertices and overall as well.
We believe that changing to flat modularity, which should also simplify
implementations, could improve clustering performance in the many
real-world processes that rely on modularity.
\end{abstract}
\maketitle

\section{Introduction}

Finding communities in graphs continues to be an important
problem. Many algorithms for doing so rely on
\textit{modularity}~\cite{eval} as the measure of the quality of the
clustering. Modularity is known to have drawbacks and other quality
measures have been proposed. Despite this, modularity continues to be
used and there is continued value in improving our understanding of
the metric.

Many graphs have degree distributions with a heavy tail. Whether these
arise from a power-law~\cite{barabasi} or some other
distribution~\cite{naysayers}, these graphs have some vertices with
very large degree and many vertices with a low degree.
Modularity explicitly treats vertices of different degrees
differently, imposing a larger penalty when high-degree vertices are
put in the same cluster. We posit that clustering algorithms that use
modularity as the score will perform poorly on lower-degree vertices.

We propose a natural variant of modularity,
\textit{flat modularity}, which does not differentiate between the vertices
based on their degrees. We use the Greedy Modularity
algorithm~\cite{greedymodularity} to cluster using modularity and
also with our proposed modularity variant on LFR graphs that have degree
distributions satisfying a power law. We compare the resulting
clusters to the planted clusters and evaluate the results using
the Matthews correlation coefficient. We observe that climbing using
our variant score improves
the match to the planted clusters. We also observe that the gain in
performance seems to come by improving the clustering of low-degree
vertices, see Figure~\vref{fig:heatdeggrps_recip}.

\section{Modularity}

Given a division of the vertices of a graph into clusters,
modularity compares the fraction of edges internal to clusters to the
average value over all graphs with the same degree distribution (where
we allow multiple edges and loops). 
One of the many ways to write the modularity of a given division into 
clusters is
\begin{equation}\label{eq:mod}
Q = \frac{1}{2L} \sum_v \sum_w C_{vw} \left(A_{vw} -
\frac{k_vk_w}{2L}\right)
\end{equation} 
where $Q$ is the modularity, $L$ the number of edges in the graph,
$k_v$ the degree of the vertex $v$, and $A$ and $C$ are adjacency
matrices, $A$ for the graph and $C$ for the clusters; that is
$C_{vw} = 1$ when $v$ and $w$ are in the same cluster and $C_{vw} = 0$
otherwise.

We interpret modularity as containing a bonus $A_{vw}$ for
putting adjacent vertices into the same cluster. The penalty
$k_vk_w/(2L)$ encourages keeping vertices separate. This penalty
therefore varies among the pairs of vertices and our primary
question is whether or not this is desirable.

Modularity is the main ingredient in a number of algorithms
intended to divide the vertices of a graph into clusters. We will use
the Greedy Modularity algorithm~\cite{greedymodularity} though 
the Louvain algorithm~\cite{louvain} is known to have a better
run-time. A more recent example can be found
in~\cite{parallelmodularity}. Algorithms based on modularity continue
to be used to identify clusterings~\cite{recent}. We chose Greedy
Modularity because it is a simple climb and we believe that this
places more weight on the score being used than a more complex climb.

It is now known~\cite{resolutionlimit} that modularity has a
fundamental drawback, called the \textit{resolution limit}.
Modularity tends to want to collapse clusters that
are smaller than this resolution limit. A common workaround is to add
a \emph{resolution} $r$ to the modularity, forming
\begin{equation}\label{eq:Q}
Q_r = \frac{1}{2L} \sum_v \sum_w C_{vw} \left(r\cdot A_{vw} -
\frac{k_vk_w}{2L}\right)
\end{equation}
for a value $r$ between 0 and 1; we reduce the reward for putting
adjacent vertices into the same cluster. This increases the relative
effect of the penalty and hence discourages combining communities.

A large number of other measures of the quality of a division into
clusters have been proposed. Many are considered
in~\cite{compscores} which also suggests different ways to evaluate
these measures. We are going to use artifical graphs with planted
clusters and that simplifies the evaluation phase.

\section{Some power-law graphs}

The penalty term in the formula for modularity, (\ref{eq:Q}),
explicitly depends on the degrees of the vertices. In many real world
graphs, for example those arising from power grids or cell
interconnections, geometry forces all vertices to have degrees in a
relatively small range. In this case, having the penalty depend on the
degree will have only a minor effect. Other real world
graphs, for example those arising from web page links or social
networks, the degree distribution can be quite heavy-tailed
(see~\cite{barabasi} for further discussion). In this
case, the magnitude of the penalty can cover a large range.

To study the consequences of the widely varying penalty, we will
use LFR graphs,~\cite{LFR}. Specifically, we use the implementation of
these graphs in the NetworkX~\cite{nx}
function {\tt LFR\_benchmark\_graph}. We will use {\tt gamma} (that is, 
$\gamma$) for the exponent in the power law of the degree distribution 
and vary a random seed {\tt seed} as we construct the graph. All of the other 
parameters are specified in Figure~\vref{fig:LFRcommand}. While one of 
our inputs seems to say that the maximum degree will be~50, in fact
our graphs commonly  have higher degrees. For one example we looked
at, the degrees ran from~14 to~67. The input parameter with value~1000
represents the number of vertices in the graph. The input parameter
with value~2 is the exponent in a power law for the community
sizes. The parameter with value~0.5 is the mixing parameter; for each
vertex roughly half of the edges will be within the planted cluster
and half will be external. We believe that this setting means that
recovering the planted clusters for the resulting graph is difficult.

\begin{figure}

\begin{center}
\begin{verbatim}
LFR_benchmark_graph(1000, gamma, 2, 0.5, average_degree=20, 
                    max_degree=50, min_community=20,
                    max_community=100, seed=seed)
\end{verbatim}
\end{center}

\caption{The command we use to generate our LFR benchmark
  graphs with inputs gamma and seed.}
\label{fig:LFRcommand}
\end{figure}

For $\gamma$ we will use $2.5$, $3.0$, and $3.5$. For each,
we will generate~1001 graphs using the seeds~0 through~1000. Later, we
will also raise the mixing parameter to~0.6 to consider even harder
problems where 60\% of the edges are external to the clusters.

\section{Finding clusterings and measuring their quality}

Given an LFR graph, we wish to find a division of the vertices into
clusters. We use the Greedy Modularity
algorithm~\cite{greedymodularity} to do so. This algorithm is fairly
simple: start with the clustering of all vertices in their own cluster
and, at each step, if combining any two clusters improves the
modularity, combine the two that give the greatest modularity. Ties
can arise; our implementation relies on sorting the improvements and
so should return the same answer on each run---we suspect that this
performs roughly the same as randomly breaking ties.

We chose this algorithm for its simplicity. Our intention is not to
measure the capability of the algorithm but rather the score on which
it is climbing. We hope that using a simple algorithm will emphasize the
differences in the quality of the score. We assume that more involved
algorithms, for example the Louvain algorithm~\cite{louvain}, will
also perform better given a better score.

Along with the authors of~\cite{darpa}, we believe that the best way
to evaluate our found clusters is to look at whether individual pairs are
correctly clustered together or apart. To be explicit, each pair of
vertices has a true label, clustered together or not, from the planted
clusters as well as a learned label from the clusters we produced via
Greedy Modularity. The $\binom{1000}{2}$ pairs
of vertices are mostly not in the same cluster, so we have unbalanced
class labels. Following the recommendation in~\cite{chooseMCC} and
elsewhere, we use the Matthews correlation coefficient~\cite{Mcc} to
turn the resulting $2\times2$ confusion matrix into a single number.
We rely on the implementation of the Matthews correlation coefficient in
Scikit-learn~\cite{sklearn}.

\section{Clustering using modularity}

\begin{figure}

\includegraphics{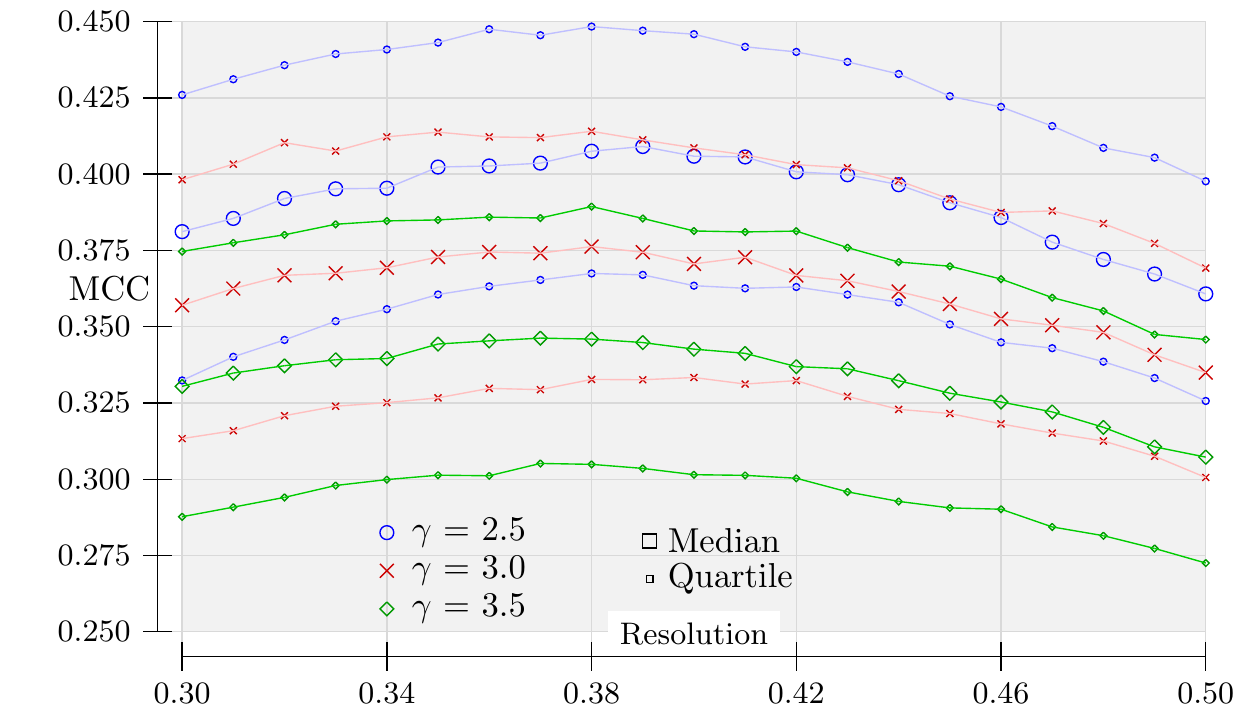}

\caption{The Matthews correlation coefficient for the clustering
  found by the Greedy Modularity algorithm using standard modularity
  $Q_r$ (\ref{eq:Q}) with resolution $r$. We present the median and
  quartile values of the MCC over 1001 LFR graphs with 1000 vertices
  and degree distributions with exponent $\gamma$; larger MCC is better.}
\label{fig:allbestres}
\end{figure}

We begin by identifying the right resolution to use for our LFR
graphs, Figure~\vref{fig:allbestres}. We have plotted the median and
quartiles of the Matthews correlation coefficient between the found
clustering and the ground truth clustering; because we used 1001
graphs, these represent the results from particular graphs. Graphs
with larger $\gamma$ prove to be harder to cluster; we contend that
this is because they contain a larger proportion of low degree vertices.
We use the greatest median to identify the best choice of
resolution, which is $0.39$ for
$\gamma = 2.5$,  $0.38$ for $\gamma = 3.0$, and $0.37$ for
$\gamma = 3.5$.

%% \begin{figure}
%% \includegraphics[width=\textwidth]{heatdeggrps.pdf}
%% \caption{Matthews correlation coefficient between the found
%%   clustering and the ground truth clustering as we restrict to
%%   vertices of different degrees for three LFR graphs with $\gamma = 2.5$.}
%% \label{fig:heatdeggrps}
%% \end{figure}

\begin{figure}
\includegraphics{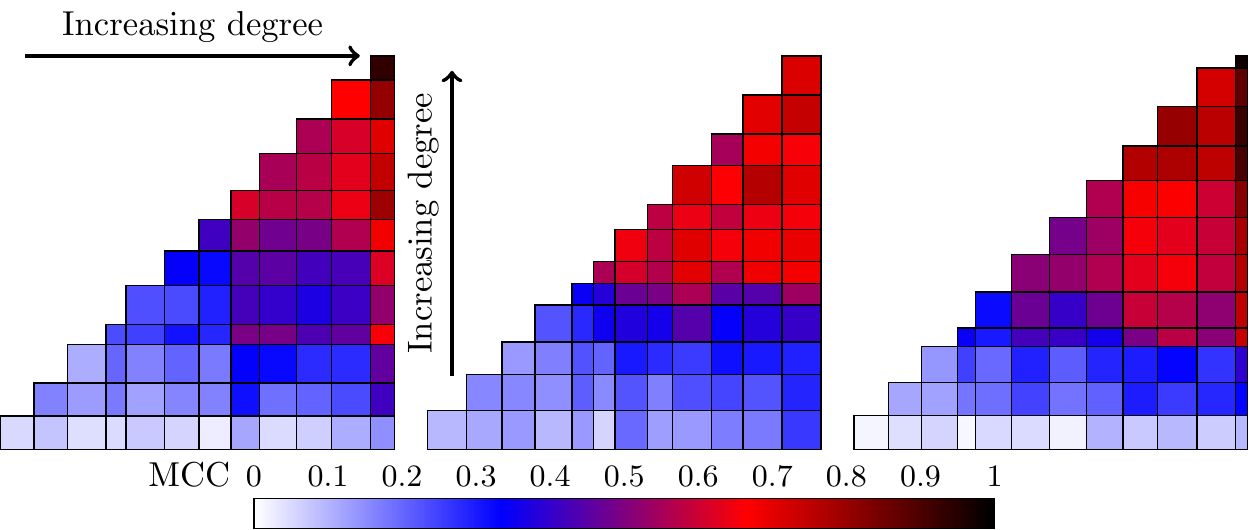}\vspace{1pt}
\includegraphics{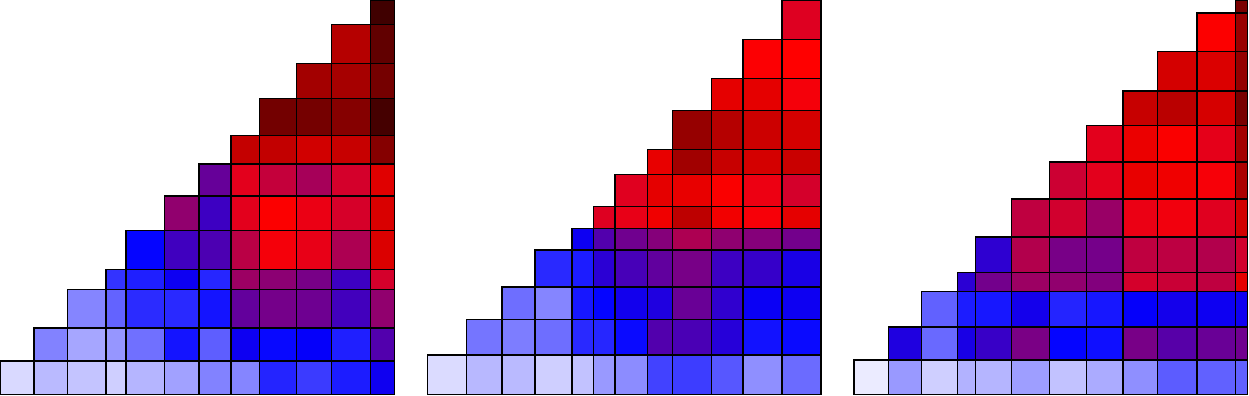}
\caption{Comparing Matthews correlation coefficient on
  different degree ranges. From left to right we have results for three LFR graphs
  with $\gamma = 2.5$; the top row represents the clustering found
  using modularity ($Q_r$ of \eqref{eq:Q} with $r = 0.39$) and the
  bottom row the clustering found using flat modularity ($Q^\flat_R$
  of \eqref{eq:flat} with $R = 98$). Within a figure, each block
  represents the Matthews correlation coefficient of the found
  clustering to the true clustering when we restrict to pairs of
  vertices with given degrees. The lowest row represents pairs
  involving vertices with the smallest degrees and the rightmost
  column represents pairs involving vertices with the largest
  degrees. A larger MCC is better.}
\label{fig:heatdeggrps_recip}
\end{figure}

We looked further at the three graphs that represent the median and
quartile scores for $r = 0.39$ with $\gamma =
2.5$. In the top row of Figure~\vref{fig:heatdeggrps_recip} we plot the
Matthews correlation coefficient between the true clusters and the
recovered clusters when we restrict to pairs of vertices with certain
degrees. For the leftmost figure, the bottom row represents vertices
of degrees 14 through 18, of which there are 85. The lower left square
represents all pairs of vertices with those degrees. The final box as
we slide to the right in that triangle
represents pairs where one vertex has degree 14 through 18 and the
other degree 48 through 61; there are only 61 such vertices, so the block
is narrower. Degrees are divided from smallest to largest with a break
when the size of the current block would exceed 100 if we included the
next degree. The next row up represents pairs involving vertices with
degrees 19 or 20.

Looking at these results, a reasonable interpretation is that when both
of the vertices have a relatively high degree, we can do pretty well at
determining whether the two vertices should be clustered together or
not. However, when at least one of the vertices has low degree, we do
a poor job. Perhaps this is an inherent problem. With higher-degree
vertices, we see more of their neighbors and so have more information
about who they should cluster with. With lower-degree vertices, we
have less information and so clustering them is harder.

However, it's also possible that the form of modularity is
deterimental to our ability to cluster the lower-degree
vertices. Making a change to modularity might be a way to improve
the performance on lower-degree vertices. We attempt this next.

\section{Flat modularity}

For modularity, the penalty term arises from averaging the
contribution over all graphs (which may have multiple edges or loops)
on this set of vertices for which the vertices have their observed
degrees. One can also interpret this as the expected value if we
randomly reconnect the half-edges emerging from each vertex.

We proceed by averaging over a larger class of graphs, all graphs on
the same vertices with the same number of edges. This changes the
expected value, resulting in the formula
\begin{equation*}
\frac{1}{2L} \sum_v \sum_w C_{vw} \left(A_{vw} -
\frac{\widehat{k}\widehat{k}}{2L}\right)
\end{equation*}
where $\widehat{k}$ is the average degree of the vertices of the
graph. When we compare with the usual modularity~(\ref{eq:mod}), we
see that the penalty term is flat (or uniform) over all pairs of
vertices and so we call this \textit{flat modularity}. Our hope is
that by not having the penalty focus on pairs of high-degree vertices,
we can achieve better performance in clustering the low-degree
vertices; we expect the additional information about high-degree
vertices to sustain our ability to cluster them well.

It's natural to think that we would want to add a resolution to flat
modularity as well. This becomes
\begin{equation*}
\frac{1}{2L} \sum_v \sum_w C_{vw} \left(r \cdot A_{vw} -
\frac{\widehat{k}\widehat{k}}{2L}\right).
\end{equation*}
In the statistical optimization literature or the machine learning
literature, it is common to subtract a regularizer which is multiplied
by some weight, usually denoted $\lambda$, that also needs to be
learned. In order to get the weight onto the penalty side, we scale
the above formula by $1/r$, arriving at
\begin{equation*}
\frac{1}{2L} \sum_v \sum_w C_{vw} \left(A_{vw} -
\frac{\widehat{k}\widehat{k}/r}{2L}\right).
\end{equation*}

In the formula above, the numerator of the penalty is some constant
written in a complicated way. We replace it with a separate constant,
arriving at our final formula for flat modularity,
\begin{equation}\label{eq:flat}
Q^\flat_R = \frac{1}{2L} \sum_v \sum_w C_{vw} \left(A_{vw} -
\frac{R}{2L}\right)
\end{equation}
We've chosen to retain the $2L$ in the denominator of the penalty in
order that the formula more closely resemble the formula for standard
modularity. We assume this aesthetic decision will have little effect
on performance but that is an open question.

As the degrees of the vertices are no longer relevant to the score, a
clustering algorithm that relies on modularity no longer needs to
track these degrees. This may simplify some implementations, though we
note that each vertex does need to know what vertices it is adjacent to.

\begin{figure}

\includegraphics{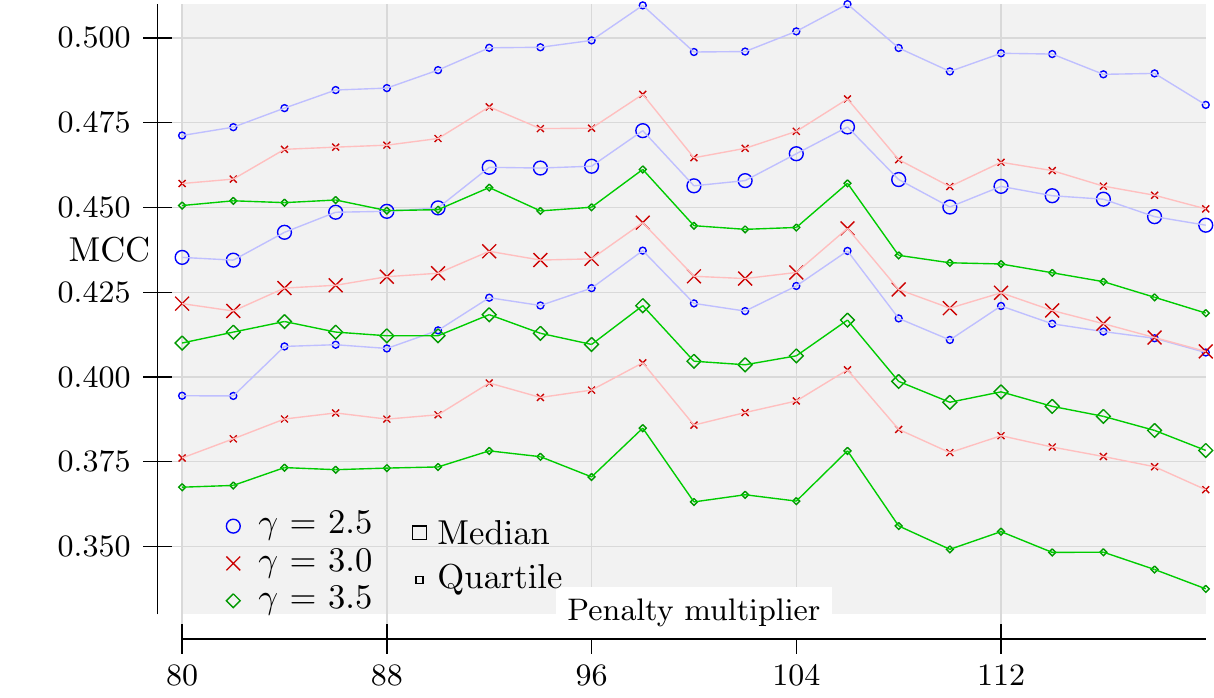}

\caption{The Matthews correlation coefficient for the clustering
  found by the Greedy Modularity algorithm using flat modularity
  $Q^\flat_R$ (\ref{eq:flat}) with penalty multiplier $R$. We present
  the median and quartile values of the MCC over 1001 LFR graphs with
  1000 vertices and degree distributions with exponent $\gamma$;
  larger MCC is better.}
\label{fig:allbestrec}
\end{figure}

For the resolution, $r$, we considered values to two decimal places
between~0 and~1. When we found that the optimal value of $R$ would be
around 100, we decided to restrict $R$ to even integers between~0
and~200. The median and quartiles of the Matthews correlation
coefficient when we apply the Greedy Modularity algorithm with flat
modularity as the score are depicted in Figure~\vref{fig:allbestrec}.

\begin{table}
\caption{Median and quartiles of the Matthews correlation
  coefficient using modularity with the best resolution we found and
  using flat modularity with the best penalty multiplier we found; a
  larger MCC is better.}
\label{tab:comp}
\begin{center}
\begin{tabular}{rrrrrrrrrrr}
\toprule
&& \multicolumn{4}{c}{Modularity} & & \multicolumn{4}{c}{Flat modularity}
  \\ %\cline{3-6} \cline{8-11}
\cmidrule{3-6} \cmidrule{8-11}
\multicolumn{1}{c}{$\gamma$} && \multicolumn{1}{c}{$r$} &
\multicolumn{1}{c}{$1/4$-ile} & \multicolumn{1}{c}{Median} &
\multicolumn{1}{c}{$3/4$-ile} && \multicolumn{1}{c}{$R$} &
\multicolumn{1}{c}{$1/4$-ile} & \multicolumn{1}{c}{Median} &
\multicolumn{1}{c}{$3/4$-ile} \\ %\cline{1-1} \cline{3-6} \cline{8-11}
\midrule
$2.5$ && $0.39$ & $0.3670$ & $0.4091$ & $0.4470$ && 98 & $0.4373$ &
$0.4727$ & $0.5096$ \\
$3.0$ && $0.38$ & $0.3327$ & $0.3762$ & $0.4140$ && 98 & $0.4042$ &
$0.4455$ & $0.4834$ \\
$3.5$ && $0.37$ & $0.3052$ & $0.3462$ & $0.3856$ && 98 & $0.3849$ &
$0.4210$ & $0.4612$ \\ \bottomrule
\end{tabular}
\end{center}
\end{table}

 Among other features, it is still the case that increasing $\gamma$
hurts our ability to recover the true clustering. The scores are also
less smooth as a function of the penalty multiplier. This might be
because there are more ties now; for modularity, the varying penalty is
likely to drive down the number of ties. We collect the best scores we
found in Table~\vref{tab:comp}. Using flat modularity has improved the
fidelity to the planted clustering and the relative decrease in
performance with rising $\gamma$ is less when we use flat modularity.

The second row of Figure~\vref{fig:heatdeggrps_recip} depicts the
results of clustering using flat modularity on the same three graphs
whose results in clustering using modularity are depicted in the first row.
Switching to flat modularity does
not hurt the performance in clustering high-degree vertices but does
improve clustering of low-degree vertices. This is particularly
evident in the lower right of the triangles which have
become much darker. Therefore, we have done a better job matching
low-degree vertices with high-degree vertices. The lower lefts of the
triangles indicate there is still room for improvement in properly
matching low-degree vertices amongst themselves, though we note that
this is likely to be difficult in general.

\begin{table}
\caption{Median and quartiles of the Matthews correlation
  coefficient over pairs where one vertex has degree at most 20 and one
  degree at least 40 split between climbing using modularity with the
  best resolution we found and using flat modularity with the best
  penalty multiplier we found; a larger MCC is better.}
\label{tab:comp2040}
\begin{center}
\begin{tabular}{rrrrrrrrrrr}
\toprule
&& \multicolumn{4}{c}{Modularity} & & \multicolumn{4}{c}{Flat modularity}
  \\ %\cline{3-6} \cline{8-11}
\cmidrule{3-6} \cmidrule{8-11}
\multicolumn{1}{c}{$\gamma$} && \multicolumn{1}{c}{$r$} &
\multicolumn{1}{c}{$1/4$-ile} & \multicolumn{1}{c}{Median} &
\multicolumn{1}{c}{$3/4$-ile} && \multicolumn{1}{c}{$R$} &
\multicolumn{1}{c}{$1/4$-ile} & \multicolumn{1}{c}{Median} &
\multicolumn{1}{c}{$3/4$-ile} \\ %\cline{1-1} \cline{3-6} \cline{8-11}
\midrule
$2.5$ && $0.39$ & $0.1705$ & $0.2115$ & $0.2538$ && 98 & $0.2243$ &
$0.2661$ & $0.3142$ \\
$3.0$ && $0.38$ & $0.1774$ & $0.2154$ & $0.2607$ && 98 & $0.2283$ &
$0.2702$ & $0.3189$ \\
$3.5$ && $0.37$ & $0.1798$ & $0.2224$ & $0.2647$ && 98 & $0.2266$ &
$0.2701$ & $0.3216$ \\ \bottomrule
\end{tabular}
\end{center}
\end{table}

These observations generally hold for all of the graphs. We computed
the Matthews correlation coefficient between the found clustering and
the true clustering restricted to pairs of vertices where one has
degree at most 20 and one degree at least 40. This corresponds to the
lower right of our triangles. The medians and quartiles over the 1001
graphs are found in Table~\vref{tab:comp2040}; the $r$ and $R$ values
are the ones from Table~\ref{tab:comp}, we did not attempt to find the
best $r$ and $R$ for this set of vertices. This supports the claim
that using flat modularity leads to improved clustering of low-degree
vertices. Note that the observed values for these pairs are more
similar for the different values of $\gamma$ than what we observed in
Table~\ref{tab:comp}. We posit that there is relatively consistent
performance among pairs of high-degree vertices, among pairs of
low-degree vertices, and among pairs where one is low-degree and the
other high-degree. Therefore, the change in performance with different
$\gamma$'s is due to the change in the numbers of each type of pair.

\begin{figure}

\includegraphics{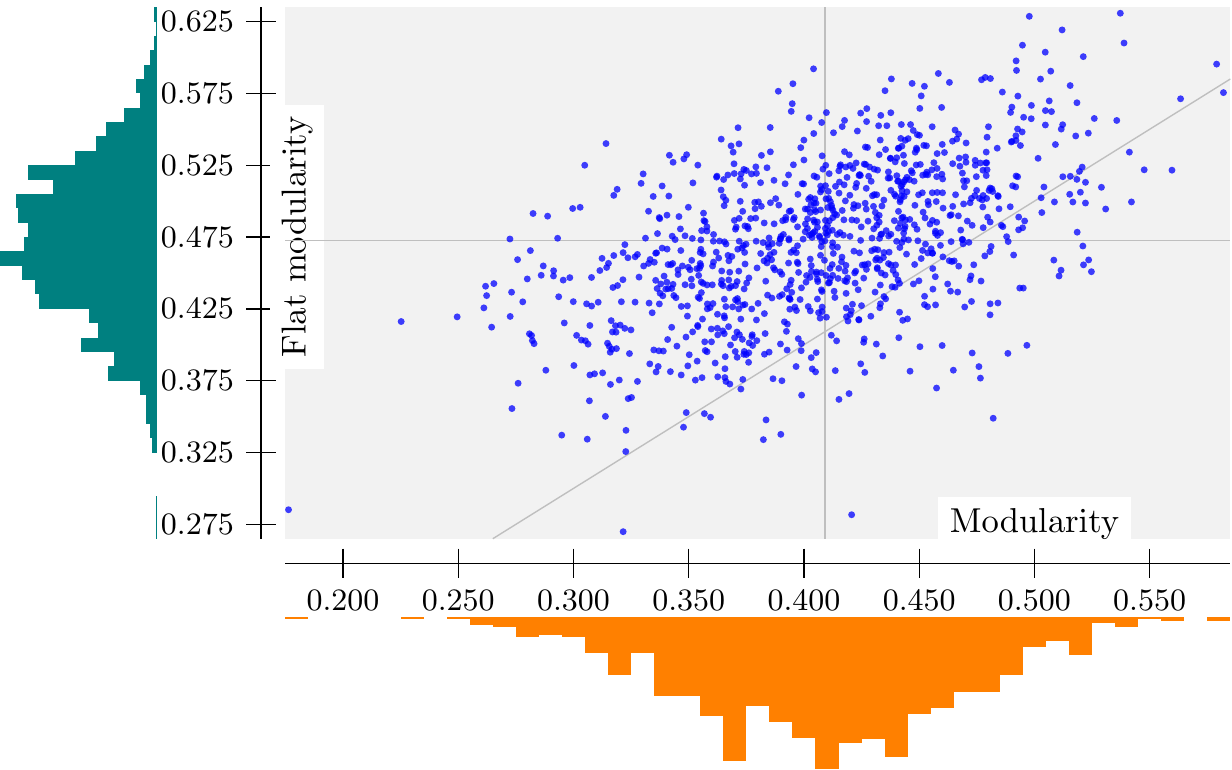}

\caption{For the 1001 LFR graphs we considered with $\gamma
  = 2.5$, we plot the Matthews correlation coefficient for the
  clustering found using modularity (x-axis; $Q_r$ of (\ref{eq:Q})
  with $r = 0.39$) against the Matthews correlation coefficient for
  the clustering found using flat modularity (y-axis; $Q^\flat_R$ of
  (\ref{eq:flat}) with $R = 98$). A larger MCC is better. The faint
  horizontal and vertical lines represent the medians of the marginal
  distributions and the faint diagonal line is $y=x$.}
\label{fig:scattercomp}
\end{figure}

 For each of the 1001 graphs with $\gamma = 2.5$, we have plotted
the Matthews correlation coefficient of the found clustering using
standard modularity with resolution $r = 0.39$ against the Matthews
correlation coefficient from using flat modularity with
multiplier $R = 98$ in Figure~\vref{fig:scattercomp}. We do not see,
for example, that the improvement only comes from graphs where the
original climb did particularly poorly; rather there is general
improvement. We have not done any analysis to identify features of the
graphs that might correlate with either score or the difference in the
scores.

We repeated these experiments with the mixing
  parameter set to %$0.4$ and 
$0.6$ instead of the $0.5$ of the main
  text. We record the results that correspond to
  Tables~\ref{tab:comp} and~\ref{tab:comp2040} in
  Table~\vref{tab:varmu}. As in the $0.5$ case, we chose $r$ and $R$
  to maximize the median Matthews correlation coefficient between the
  planted clusters and the found clusters over all of the vertices
  and, for those values of $r$ and $R$, we report out statistics for
  the Matthews correlation coefficient between the planted clustering
  and the found clustering restricted to pairs of vertices where one
  vertex has degree at most 20 and one degree at least 40.
In this harder case, there is again a large difference in favor of flat
modularity in both the overall scores and restricted to pairs that involve one
low-degree vertex and one high-degree vertex.

\begin{table}
\caption{Median and quartiles of the Matthews correlation
  coefficient for either all pairs of vertices or only pairs where one
  vertex has degree at most 20 and one degree at least 40 split between
  climbing using modularity with the best resolution we found and
  using flat modularity with the best penalty multiplier we found over 1001
  %graphs where the mixing parameter was either $0.4$ or $0.6$; a
  graphs where the mixing parameter was $0.6$; a
  larger MCC is better. The chosen values of $r$ and $R$ maximize the
  median Matthews correlation over all pairs of vertices.}
\label{tab:varmu}
\begin{center}
\begin{tabular}{rrrrrrrrrrr}\toprule
&& \multicolumn{4}{c}{Modularity} & & \multicolumn{4}{c}{Flat modularity}
  \\ %\cline{3-6} \cline{8-11}
\cmidrule{3-6} \cmidrule{8-11}
\multicolumn{1}{c}{$\gamma$} && \multicolumn{1}{c}{$r$} &
\multicolumn{1}{c}{$1/4$-ile} & \multicolumn{1}{c}{Median} &
\multicolumn{1}{c}{$3/4$-ile} && \multicolumn{1}{c}{$R$} &
\multicolumn{1}{c}{$1/4$-ile} & \multicolumn{1}{c}{Median} &
\multicolumn{1}{c}{$3/4$-ile} \\ %\cline{1-1} \cline{3-6} \cline{8-11}
\midrule
%% && \multicolumn{9}{c}{Mixing parameter $0.4$, all pairs} \\
%% $2.5$ && $0.33$ & $0.7556$ & $0.7841$ & $0.8098$
%%           && 92 & $0.7588$ & $0.7921$ & $0.8189$ \\
%% $3.0$ && $0.33$ & $0.7374$ & $0.7680$ & $0.7931$
%%           && 92 & $0.7341$ & $0.7681$ & $0.8010$ \\
%% $3.5$ && $0.32$ & $0.7129$ & $0.7466$ & $0.7769$
%%           && 84 & $0.7110$ & $0.7462$ & $0.7806$ \\
%% && \multicolumn{9}{c}{Mixing parameter $0.4$, pairs with low degree and
%%   high degree} \\
%% $2.5$ && $0.33$ & $0.6148$ & $0.6645$ & $0.7062$
%%           && 92 & $0.5952$ & $0.6462$ & $0.6951$ \\
%% $3.0$ && $0.33$ & $0.5944$ & $0.6485$ & $0.6875$ 
%%           && 92 & $0.5872$ & $0.6478$ & $0.6954$ \\
%% $3.5$ && $0.32$ & $0.6207$ & $0.6728$ & $0.7155$
%%           && 84 & $0.5832$ & $0.6366$ & $0.6884$ \\
%% && \multicolumn{9}{c}{Mixing parameter $0.6$, all pairs} \\
&& \multicolumn{9}{c}{All pairs} \\ \midrule
$2.5$ && $0.39$ & $0.0792$ & $0.1072$ & $0.1322$ && 
            106 & $0.1535$ & $0.1880$ & $0.2186$ \\
$3.0$ && $0.39$ & $0.0723$ & $0.0949$ & $0.1177$ &&
            106 & $0.1386$ & $0.1700$ & $0.2032$ \\
$3.5$ && $0.44$ & $0.0630$ & $0.0861$ & $0.1060$ &&
            106 & $0.1225$ & $0.1529$ & $0.1826$ \\ \midrule
%% && \multicolumn{9}{c}{Mixing parameter $0.6$, pairs with low degree and
%%   high degree} \\
&& \multicolumn{9}{c}{Pairs with a low-degree vertex and
  a high-degree vertex only} \\ \midrule
$2.5$ && $0.39$ & $0.0188$ & $0.0271$ & $0.0383$ &&
            106 & $0.0328$ & $0.0468$ & $0.0649$ \\
$3.0$ && $0.39$ & $0.0231$ & $0.0334$ & $0.0465$ &&
            106 & $0.0379$ & $0.0535$ & $0.0728$ \\
$3.5$ && $0.44$ & $0.0263$ & $0.0363$ & $0.0484$ &&
            106 & $0.0422$ & $0.0588$ & $0.0804$ \\ \bottomrule
\end{tabular}
\end{center}
\end{table}

\section{Conclusions}

 Modularity continues to be a factor in finding clusters in
graphs. If we interpret modularity as involving a bonus and a penalty,
then the penalty is higher for higher-degree vertices. When we use a simple
clustering algorithm that attempts to maximize modularity, it fares
far better on the high-degree vertices than on the low-degree vertices.

 We therefore ask whether it is better to have the
penalty be independent of the degree of the vertex. We found that the same
simple algorithm, now using flat modularity, finds clusterings
that better match the ground truth, in particular improving the
clustering of lower-degree vertices. Meanwhile, our implementation
became simpler because we no longer needed to track the degrees of the
vertices.

 We have more information about the vertices with higher degrees and so
it is not surprising that we can cluster them more
effectively. Having our penalty focus on this high information region
proves less helpful than flattening the penalty.
Perhaps one could even go further and try to apply
the penalty disproportionately to low-degree vertices; we have not
tested this idea.

 While our tests were confined to graphs in which the vertex degrees
satisfy a power law, we expect the results to generalize to other
heavy-tailed degree distributions. Since such graphs occur frequently, and
finding communities in graphs continues to be an important problem, we
believe that the simple change from modularity to flat modularity can
improve a broad range of clustering results with minimal change to
implementations.

\end{document}